\documentclass[a4paper]{jpconf}
\usepackage{graphicx}
\usepackage{subfigure}
\usepackage[capitalise]{cleveref}
\usepackage{lineno}
\newcommand{\nuebar}{$\overline{\nu}_{e}$}

\begin{document}
\title{JUNO Oscillation Physics}

\author{Jinnan Zhang$^{1,2}$ on behalf of the JUNO collaboration}

\address{$^1$ Institute~of~High~Energy~Physics, Beijing, China}
\address{$^2$ University of Chinese Academy of Sciences, Beijing, China}

\ead{zhangjinnan@ihep.ac.cn}

\begin{abstract}
    The Jiangmen Underground Neutrino Observatory (JUNO) is a 20 kton multi-purpose liquid scintillator detector with an expected $3\%/\sqrt{E[\mbox{Mev}]}$ energy resolution, under construction in a 700 m underground laboratory in the south of China (Jiangmen city, Guangdong province). The exceptional energy resolution and the massive fiducial volume of the JUNO detector offer great opportunities for addressing many essential topics in neutrino and astroparticle physics. JUNO's primary goals are to determine the neutrino mass ordering and precisely measure the related neutrino oscillation parameters. With six years of data taking with reactor antineutrinos, JUNO can determine the mass ordering at a 3-4$\sigma$ significance and the neutrino oscillation parameters $\sin^2\theta_{12}$, $\Delta m^2_{21}$, and $|\Delta m^2_{31}|$ to a precision of better than 0.6\%. In addition, the atmospheric neutrino and solar neutrino measurement at JUNO can also provide complementary and important information for neutrino oscillation physics. This work focuses on the oscillation physics of JUNO, which includes measurement and analysis of the reactor neutrinos, the atmospheric neutrinos, and the solar neutrinos. With the delicate energy response calibration and event reconstruction potential, JUNO will make a world-leading measurement on the neutrino oscillation parameters and neutrino mass ordering in the near future.
\end{abstract}

\section{Introduction}
\label{sec:Introduction}
The standard three-flavor neutrino oscillation pattern with the Pontecorvo-Maki-Nakagawa-Sakata (PMNS) matrix describing the mixing between flavor and mass eigenstates is well established and verified by plenty of neutrino experiments. There are many open questions in neutrino oscillation physics; the neutrino mass ordering (NMO) is one of the main focuses of the next step neutrino oscillation experiments, i.e., to determine whether the 3rd generation $\nu_3$ is heavier or lighter than the first two generations $\nu_1$ and $\nu_2$. The normal ordering (NO) refers to $m_3>m_2>m_1$ and inverted ordering (IO) refers to $m_3<m_1<m_2$. The $\nu_1$ and $\nu_3$ are neutrino mass eigenstates with largest and smallest mixture of $\nu_e$, respectively. The three neutrino mixing angles $\theta_{12}$, $\theta_{23}$, $\theta_{13}$ and the two independent mass squared splitting $\Delta m_{21}^2$ and $|\Delta m_{31}^2|$ (or $|\Delta m_{32}^2|$) have been determined with a precision of a few percent~\cite{ParticleDataGroup:2020ssz}.

\begin{figure}[h]
    \begin{minipage}{0.44\textwidth}
        \includegraphics[width=\textwidth]{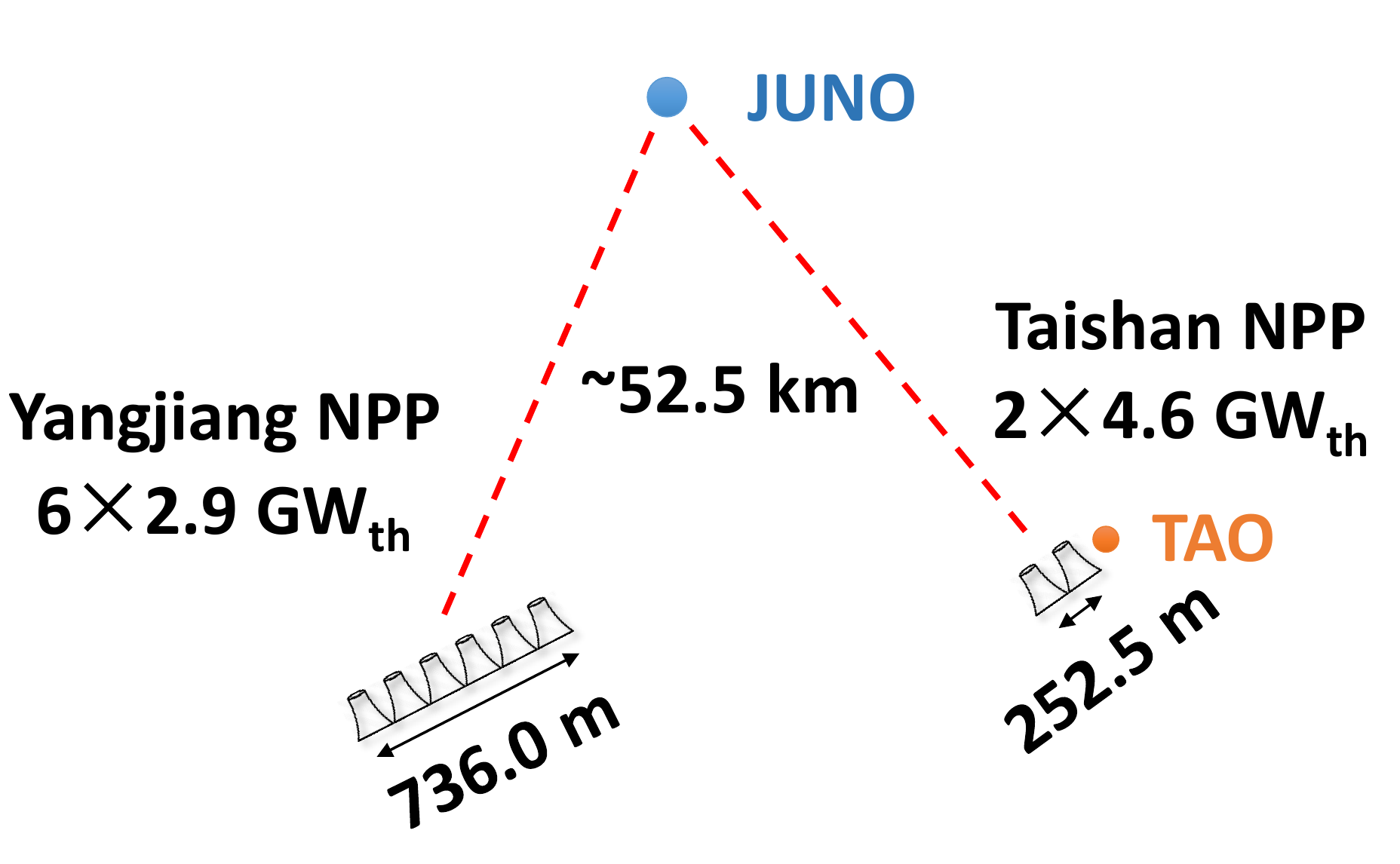}
        \caption{\label{fig:JUNO:setup}The JUNO experiment setup.}
    \end{minipage}\hspace{2pc}%
    \begin{minipage}{0.46\textwidth}
        \includegraphics[width=\textwidth]{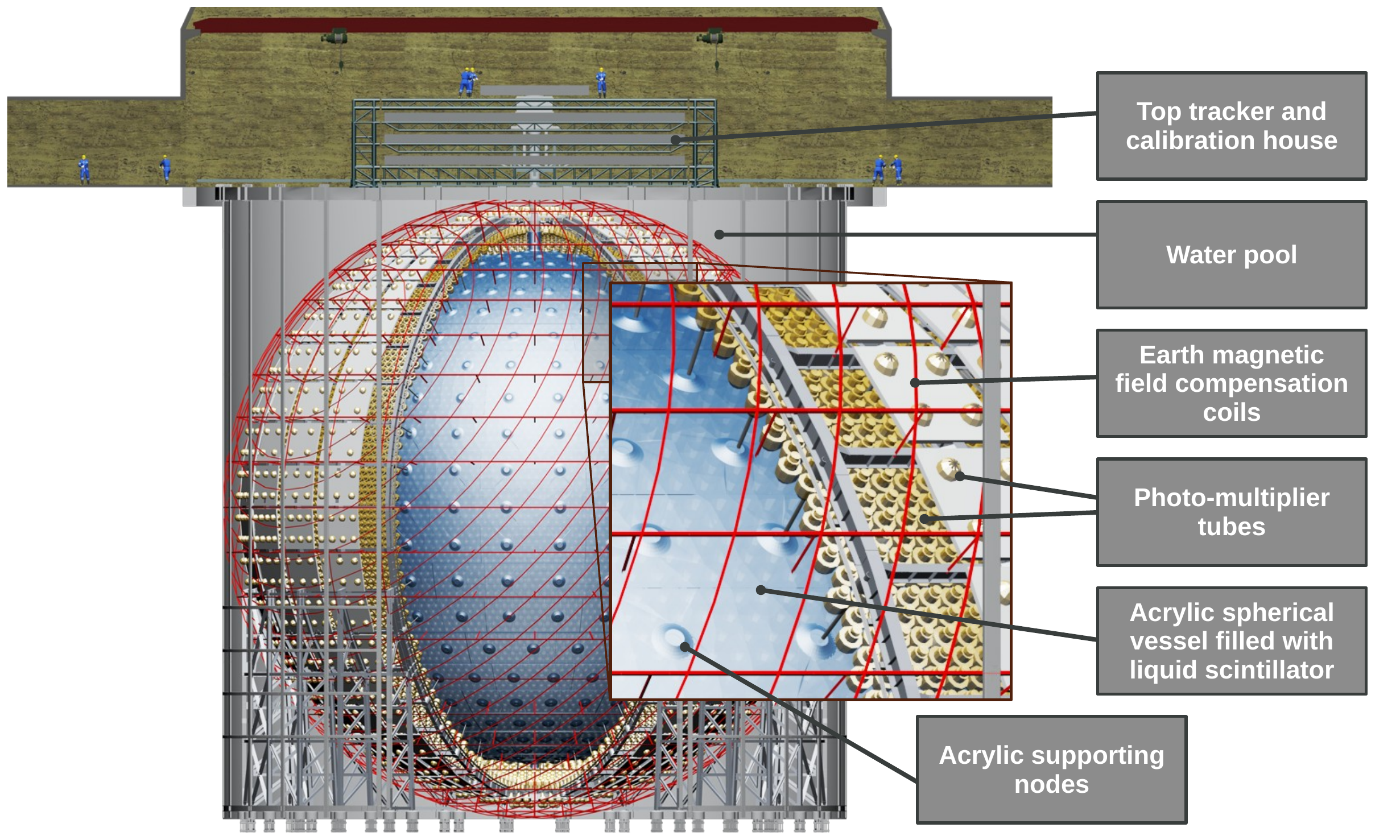}
        \caption{\label{fig:JUNO:detector}The JUNO detector schematic view. The liquid scintillator in the acrylic sphere is surrounded by 20-inch and 3-inch PMTs.}
    \end{minipage}
\end{figure}

The Jiangmen Underground Neutrino Observatory (JUNO) is a 20 kton multi-purpose liquid scintillator detector with an expected $3\%/\sqrt{E[\mbox{Mev}]}$ energy resolution, under construction in a 700 m underground laboratory in the south of China (Jiangmen city, Guangdong province). The major physics goals of JUNO are to determine the NMO and precisely measure the oscillation parameters~\cite{JUNO:2021vlw}. The location of JUNO is optimized for determining the NMO with reactor antineutrinos mainly from Taishan (2$\times$ 4.6 ${\rm GW}_{\rm th}$) and Yangjiang (6$\times$ 2.9 ${\rm GW}_{\rm th}$) Nuclear Power Plant (NPP). \cref{fig:JUNO:setup} shows the location of the JUNO experiment and the satellite experiment Taishan Antineutrino Detector (TAO~\cite{JUNO:2020ijm}), which is about 30~m from one of the Taishan reactor cores. \cref{fig:JUNO:detector} shows the schematic design of the JUNO central detector, which provides high optical coverage and large statistics for the physics goals.
This paper will introduce the JUNO NMO determination and precision measurement of oscillation parameters with various sources of neutrinos. The main contribution is from the reactor antineutrinos, while atmospheric neutrinos and solar neutrinos will provide complementary measurements.

\section{Reactor antineutrinos at JUNO}
\label{sec:ReactorNeutrinos}
The antineutrinos from the reactors close to JUNO provide the primary signal for oscillation physics at JUNO. The reactors around the JUNO experiment site are commercial light-water reactors, in which more than 99.7\% of the antineutrinos are from the fission of the four isotopes, $^{235}$U, $^{238}$U, $^{239}$Pu and $^{241}$Pu. JUNO measures the reactor electron antineutrino via inverse beta decay (IBD) interaction: $\bar{\nu}_e+p\to e^{+}+n.$ The $\bar{\nu}_e$ interacts with proton in the liquid scintillator (LS), creating a positron ($e^+$) and a neutron ($n$). The $e^+$ quickly deposits its energy and annihilates into two 0.511-MeV photons, gives a prompt signal. The neutron will be captured by hydrogen ($\sim$99\%) or carbon ($\sim$1\%) after approximately 200 $\mu s$ scattering in the detector, producing gammas of 2.22 MeV or 4.95 MeV, respectively; the capture of neutron will give a delayed signal.

\begin{figure}[h]
    \begin{minipage}{0.44\textwidth}
        \includegraphics[width=\textwidth]{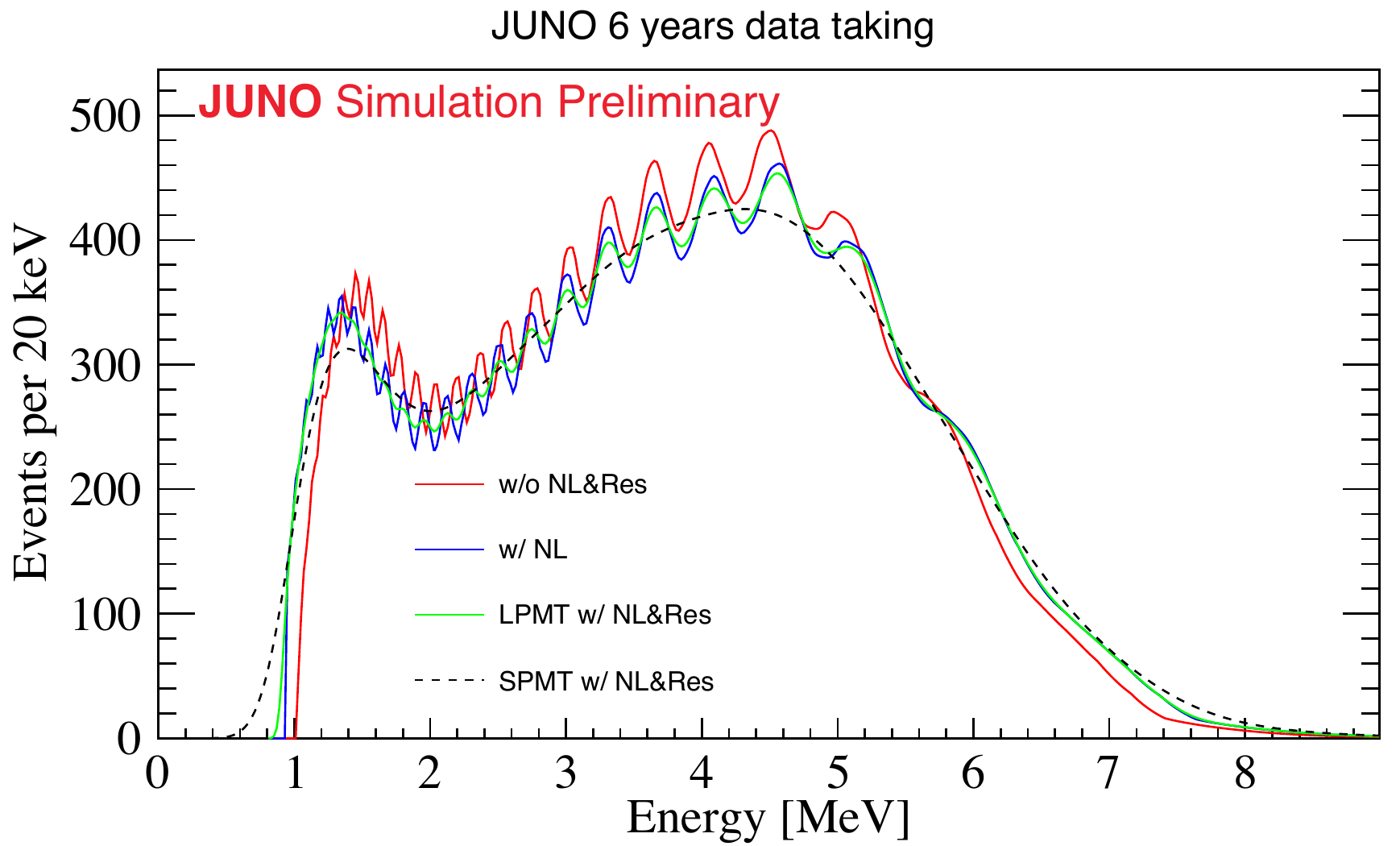}
        \caption{\label{fig:DetectorResponse}The detector response effects impact on the IBD signal spectrum.}
    \end{minipage}\hspace{2pc}%
    \begin{minipage}{0.44\textwidth}
        \includegraphics[width=\textwidth]{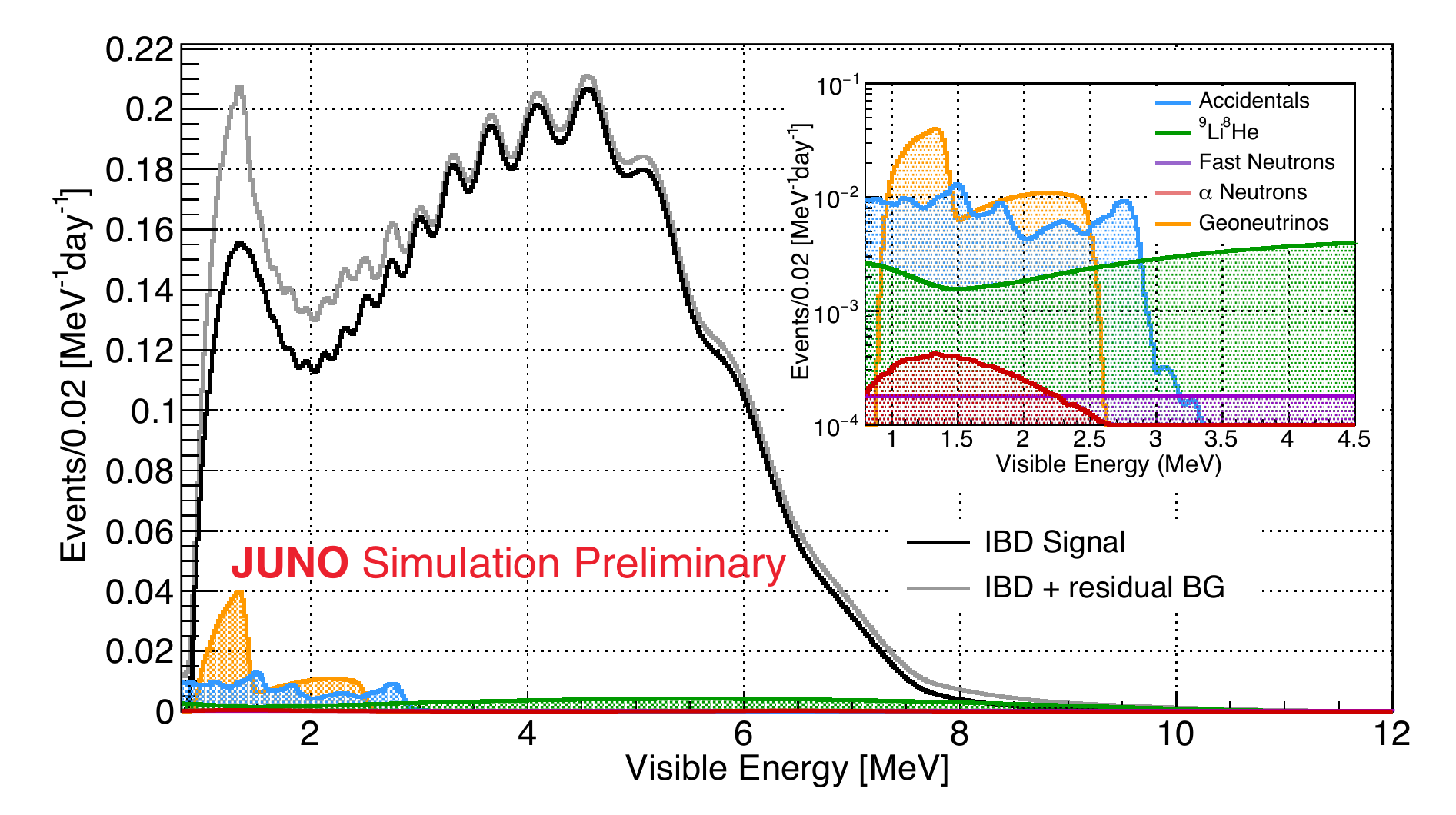}
        \caption{\label{fig:SignalBkg}The expected reactor antineutrino IBD signal and background spectra at JUNO.}
    \end{minipage}
\end{figure}

The \nuebar~transfers most of the energy to the positron via IBD interaction; thus, we extract the oscillation physics by performing analysis on the reconstructed energy of the prompt signal. The positrons deposit energy in the detector and generate photons by the scintillation and Cherenkov radiation processes. The photons are collected by the Photomultiplier Tubes (PMTs) and converted to photoelectrons (PE), and read out by the electronics. The total number of photons is not linear with respect to the total deposited energy of positron, and the nonlinear relationship between the deposited energy of positron and expected visible energy ($E^0_{\rm vis}$) is noted as the non-linearity (NL). The reconstructed visible energy $E^{\rm rec}_{\rm vis}$ in a real detector will follow the Gaussian distribution around $E^0_{\rm vis}$. Due to LS's close composition of the Daya Bay experiment and JUNO~\cite{JUNO:2020bcl}, we construct the detector response model by employing the liquid scintillator non-linearity curve of Daya Bay~\cite{DayaBay:2019fje} with proper energy scale and energy resolution given by the simulation of JUNO calibration strategy in Ref.~\cite{JUNO:2020xtj}. JUNO has two independent photon collection and readout systems with different photon occupancy regimes. One is the 17,612 20-inch large PMTs (LPMT) system, and the other is the 25,600 3-inch small PMTs (SPMT) system~\cite{JUNO:2021vlw}. The LPMT system, thanks to the excellent expected performance (high photon detection efficiency and optical coverage), will allow to reach the challenging energy resolution requirements. The SPMT system, whose energy resolution performance is worse than LPMTs, will look at the same events as the LPMTs and has the potential to measure solar oscillation parameters semi-independently. \cref{fig:DetectorResponse} shows the reactor antineutrino energy spectrum at JUNO under different detector response effects. We can see that the non-linearity effect will distort the spectrum, and the energy resolution of the LPMT system and SPMT system have different smearing on the spectrum.

The energy characteristics, time correlation, and spatial correlation between the prompt and delayed signals allow to reduce the background events significantly. Based on the measurement and simulation, we estimated the background rate and spectra for reactor antineutrino analysis as shown in \cref{fig:SignalBkg}.

\begin{figure}[h]
    \includegraphics[width=0.45\textwidth]{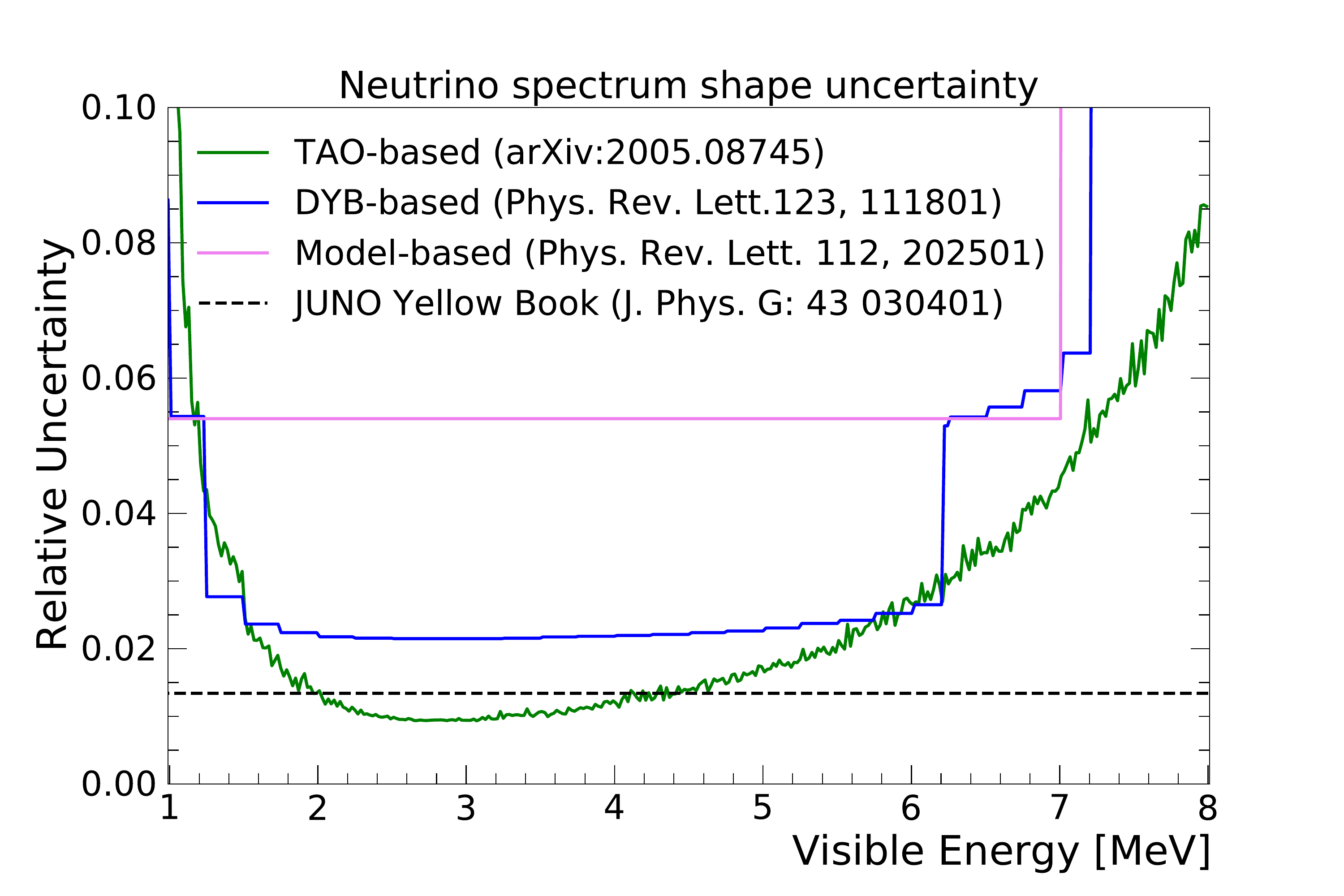}\hspace{2pc}%
    \begin{minipage}[b]{0.45\textwidth}\caption{\label{fig:UncertaintyModels} The reactor antineutrino energy spectrum relative uncertainty from different models. We display the uncertainties on observed visible energy with corresponding energy conversion of JUNO.}
    \end{minipage}
\end{figure}

Apart from the main detector JUNO, we have the satellite detector near the Taishan NPP. TAO is about 30 m from one of the Taishan reactor cores and will operate at -50 $^\circ$C with about 2\%/$\sqrt{E[{\rm MeV}]}$ energy resolution.  The initial purpose of TAO is to provide a reference for JUNO and control the shape uncertainty of the reactor antineutrino energy spectrum~\cite{JUNO:2020ijm}. With the larger statistics and good energy resolution of TAO, we can measure the spectrum to unprecedented precision. \cref{fig:UncertaintyModels} shows the power of TAO on constraining the shape uncertainty comparing to other models.

\subsection{Neutrino mass ordering determination}
To obtain the JUNO sensitivity, an Asimov data set, following the processes above under either normal ordering or inverted ordering hypothesis, was generated. A chi-square was obtained by fitting to an Asimov data with the assumptions of both normal and inverted ordering and take the difference of minima in \cref{eq:NMO:sensitivity} as a measure of NMO sensitivity~\cite{JUNO:2021vlw}.

\begin{equation}\label{eq:NMO:sensitivity}
    \Delta\chi^2_{\rm NMO}=\left|\chi^2_{\rm min}({\rm NO})-\chi^2_{\rm min}({\rm IO})\right|,
\end{equation}
where $\chi^2_{\rm min}({\rm NO})$ and $\chi^2_{\rm min}({\rm IO})$ are the minima of the chi-square function under the NO and IO hypothesis, respectively.
We find that the median sensitivity of JUNO on determining the NMO would be larger than 3$\sigma$ for 6 years of data taking.
\subsection{Precision measurement of oscillation parameters}
The precision measurement of the oscillation parameters plays a vital role in neutrino physics. For a given oscillation parameter x (would be $\Delta m^2_{21}$, $\Delta m_{31}^2$, $\sin^2{\theta_{12}}$, or $\sin^2{\theta_{13}}$.), we can define the $1\sigma$ relative precision in \cref{eq:RelativePrecision}.

\begin{equation}
    P_{1\sigma}=\frac{(x^{\rm up}-x^{\rm low})}{2\cdot x_{\rm bft}},
    \label{eq:RelativePrecision}
\end{equation}
where $x^{\rm up}$($x^{\rm low}$) is the upper (lower) bound for parameter x at $1\sigma$ level, i.e., at which the marginalized $\Delta\chi^2(x)\left(\equiv \chi^2(x)-\chi^2_{\rm min}\right)$ is equal to 1. $x_{\rm bft}$ is the best fit value of the oscillation parameters.

\begin{figure}[h]
    \includegraphics[width=0.56\textwidth]{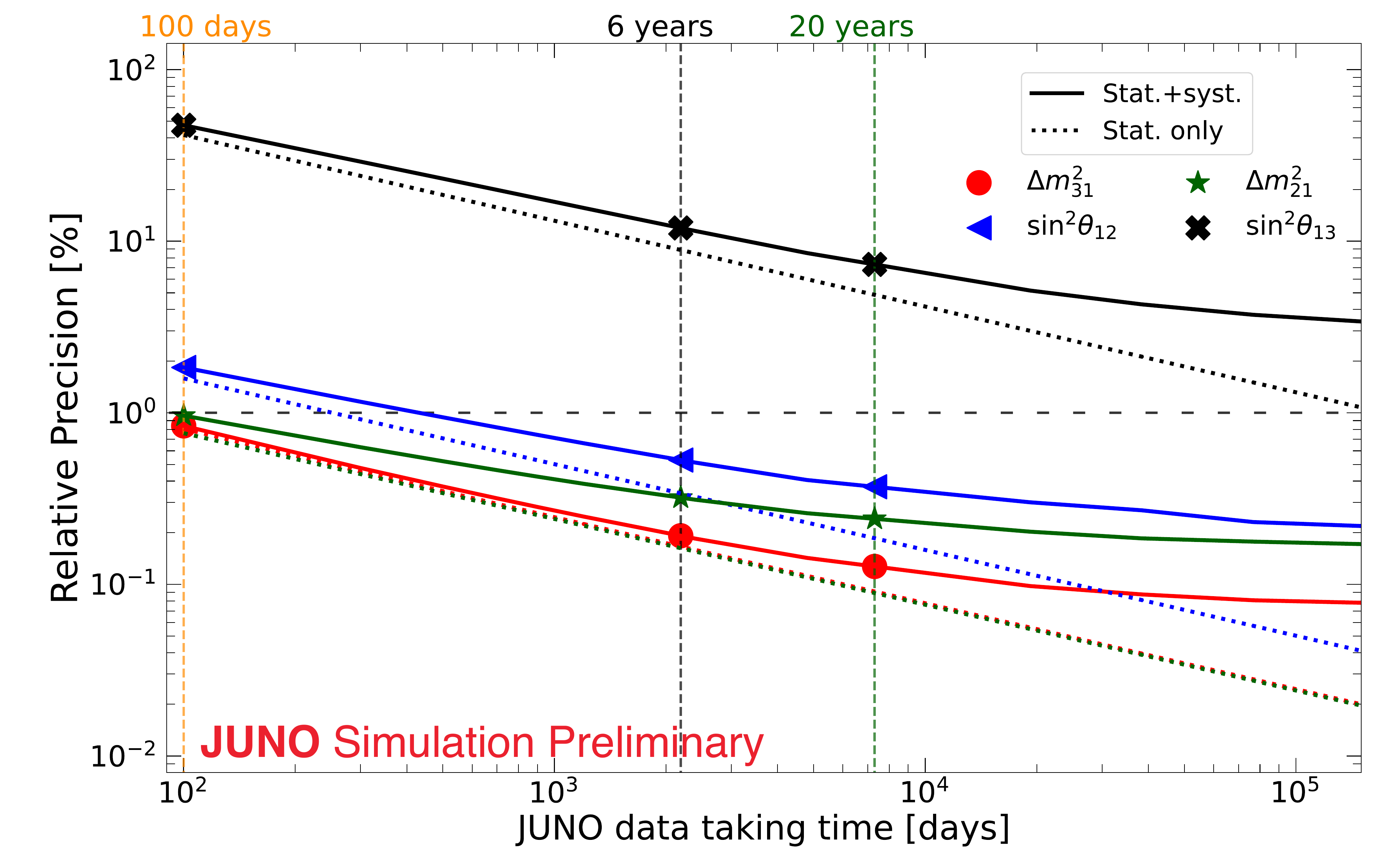}\hspace{2pc}%
    \begin{minipage}[b]{0.33\textwidth}
        \caption{\label{fig:SensitivityOverTimeBOTH}Relative precision of oscillation parameters as a function of JUNO data taking time. The dashed and solid lines are the sensitivity of considering only statistical uncertainty and all uncertainties, respectively.}
    \end{minipage}
\end{figure}
With a large number of reactor antineutrinos and excellent detector performance, JUNO will be able to measure the neutrino oscillation parameters $\sin^2{\theta_{12}}$, $\Delta m_{21}^2$ and $\Delta m_{31}^2$ to an unprecedented precision of better than 1\%. \cref{fig:SensitivityOverTimeBOTH} shows the time evolution of the JUNO precision sensitivity, which indicates the precision of oscillation parameters will come to the sub-percent era after about one year of data taking of JUNO. 
\begin{table}[htbp]
    \caption{The expected precision levels for the oscillation parameters with 6~years of run time and all uncertainties considered. The current PDG2020 sensitivity values are shown for comparison.}
    \label{tab:sensitivty}%
    \centering
    \begin{tabular}{lllll}
        \br
                                                 & $|\Delta m^{2}_{31}|$ & $\Delta m^{2}_{21}$ & $\sin^{2}\theta_{12}$ & $\sin^{2}\theta_{13}$ \\
        \mr
        JUNO 6 years                             & $\sim$0.2\%           & $\sim$0.3\%         & $\sim$0.5\%           & $\sim$12\%            \\

        PDG2020 & $ 1.4  \%$            & $ 2.4 \%$           & $ 4.2  \%$            & $ 3.2  \%$            \\
        \br
    \end{tabular}%

\end{table}
\cref{tab:sensitivty} shows the precision measurement performance of JUNO after 6 years of data taking, comparing to the current knowledge given by PDG2020~\cite{ParticleDataGroup:2020ssz}.
\section{Atmospheric and solar neutrinos}
\label{sec:AtmSolarNeutrinos}
\subsection{Atmospheric neutrinos at JUNO}
The anticipated larger target volume, energy, position, and direction reconstruction capabilities allow JUNO to explore the oscillation physics with the atmospheric neutrinos. The atmospheric neutrinos are generated by cosmic rays interacting with the Earth's atmosphere. During the propagation to JUNO, the neutrinos will oscillate and can be predicted by taking matter effect into account. At JUNO, the neutrinos will interact with nuclei in the detector via charged current (CC) or neutral current (NC) interactions~\cite{JUNO:2021tll}. By looking at the final states, we can classify the events into different samples for the oscillation physic analysis of the atmospheric neutrinos. With the event characteristics given by the detector simulation, we select the samples based on the information of the Michel electrons, neutrino capture, and unstable daughter nuclei in the final states. Then we perform the fit on classified samples of the $\bar{\nu}_\mu$, $\nu_\mu$, $\bar{\nu}_e$, and $\nu_e$ CC events. We find that JUNO can determine the NMO with atmospheric neutrinos with more than 1$\sigma$ significance for 6 years of data taking. Besides, the atmospheric neutrinos at JUNO will provide complementary measurement on determining the octant of $\theta_{23}$ and value of CP violation phase $\delta_{\rm CP}$~\cite{JUNO:2021vlw}.
\subsection{Solar neutrinos at JUNO}
The detection of the neutrinos produced by the thermonuclear fusion reactions in the solar core can allow to measure neutrino oscillation physics. With the elastic scattering interaction, JUNO can probe the ${}^8$B solar neutrino with a threshold of 2.5 MeV. After 10 years of data taking, JUNO will detect about 60k neutrino signals with approximately 30k background events. The data can provides the opportunity to measure the day-night-asymmetry with 0.9\% sensitivity, which is better than the Super-K experiment. The analysis also yields that the data can independently measure the oscillation parameters $\Delta m^2_{21}$ to 20\% precision and $\sin^2\theta_{12}$ to 8\% precision~\cite{JUNO:2020hqc}.
\section{Conclusion}
In about one year after JUNO starts operation in 2023, we can measure the oscillation parameters $\sin^2\theta_{12}$, $\Delta m^2_{21}$, and $|\Delta m^2_{31}|$ to the sub-percent precision with the reactor antineutrinos. After about 6 years of data taking, JUNO can determine the neutrino mass ordering with larger than 3$\sigma$ significance. We can combine the reactor antineutrino data with the atmospheric neutrino data; in this way, the JUNO NMO determination significance can be improved by more than 1$\sigma$. The atmospheric neutrinos and solar neutrinos will also provide the independent and complementary measurement of the oscillation parameters, i.e., $\theta_{23}$ octant and $\delta_{\rm CP}$ measurement by atmospheric neutrinos data, and $\Delta m^2_{21}$ and $\sin^2{\theta_{12}}$ measurement by solar neutrino data.

\section*{References}
\bibliography{TAUP2021-JPCS-Proceeding}

\providecommand{\newblock}{}
\begin{thebibliography}{1}
\expandafter\ifx\csname url\endcsname\relax
  \def\url#1{{\tt #1}}\fi
\expandafter\ifx\csname urlprefix\endcsname\relax\def\urlprefix{URL }\fi
\providecommand{\eprint}[2][]{\url{#2}}

\bibitem{ParticleDataGroup:2020ssz}
Zyla P~A {\em et~al.\/} (Particle Data Group) 2020 {\em PTEP\/} {\bf 2020}
  083C01

\bibitem{JUNO:2021vlw}
Abusleme A {\em et~al.\/} (JUNO) 2021  (\textit{Preprint} \eprint{2104.02565})

\bibitem{JUNO:2020ijm}
Abusleme A {\em et~al.\/} (JUNO) 2020  (\textit{Preprint} \eprint{2005.08745})

\bibitem{JUNO:2020bcl}
Abusleme A {\em et~al.\/} (JUNO, Daya Bay) 2021 {\em Nucl. Instrum. Meth. A\/}
  {\bf 988} 164823 (\textit{Preprint} \eprint{2007.00314})

\bibitem{DayaBay:2019fje}
Adey D {\em et~al.\/} (Daya Bay) 2019 {\em Nucl. Instrum. Meth. A\/} {\bf 940}
  230--242 (\textit{Preprint} \eprint{1902.08241})

\bibitem{JUNO:2020xtj}
Abusleme A {\em et~al.\/} (JUNO) 2021 {\em JHEP\/} {\bf 03} 004
  (\textit{Preprint} \eprint{2011.06405})

\bibitem{JUNO:2021tll}
Abusleme A {\em et~al.\/} (JUNO) 2021 {\em Eur. Phys. J. C\/} {\bf 81} 10
  (\textit{Preprint} \eprint{2103.09908})

\bibitem{JUNO:2020hqc}
Abusleme A {\em et~al.\/} (JUNO) 2021 {\em Chin. Phys. C\/} {\bf 45} 023004
  (\textit{Preprint} \eprint{2006.11760})

\end{thebibliography}
\end{document}